\documentclass[aps,prl,twocolumn,10pt,superscriptaddress,nofootinbib]{revtex4-2}

\usepackage{amsmath,amssymb,mathtools}
\usepackage{bm}
\usepackage{comment}

\makeatletter
\def\label#1{\@bsphack
  \begingroup
  \UseHookWithArguments{label}{1}{#1}%
  \protected@write\@auxout{}%
    {\string\newlabel{#1}{{\@currentlabel}{\thepage}%
      {\@currentlabelname}{\@currentHref}{\@kernel@reserved@label@data}}}%
  \endgroup
  \@esphack}
\let\ltx@label\label
\makeatother
\usepackage{hyperref}

\makeatletter
\patchcmd{\@bibdataout@aps}{author="08"}{author="48"}{}{}
\patchcmd{\@bibdataout@aps}{author="08"}{author="48"}{}{}
\makeatother

\newcommand{\cI}{\mathcal I_2}
\newcommand{\OO}{\mathcal O}
\newcommand{\bbone}{\mathbf 1}
\newcommand{\Tr}{\operatorname{Tr}}

\begin{document}

\title{Birth and Death in Two-color ABJM}

\author{Marco S. Bianchi} \affiliation{Facultad de Ingeniería, Universidad San Sebastián, Santiago, Chile}

\email{marco.bianchi@uss.cl}

%\date{\today}

\begin{abstract}
We solve exactly the leading quantum deformation of the two-point metric of half-BPS operators in $U(2)\times U(2)$ ABJM theory at finite rank.  In the natural coupling-independent Schur frame, its tree-normalized two-loop correction is a Jacobi operator on the chain of two-row Young diagrams.  A ground-state transform maps it to a reversible birth--death process whose stationary distribution is the Plancherel measure conditioned on diagrams with at most two rows.  Its eigenmodes are Racah polynomials, its large-$n$ limit approaches radial Ornstein--Uhlenbeck dynamics, and an equivalent two-spin description identifies the Jacobi operator with a restricted $SU(2)$ Casimir. 
\end{abstract}

\maketitle

\section{Introduction}
ABJM theory is a three-dimensional $\mathcal N=6$ superconformal Chern--Simons-matter theory with gauge group $U(N)_\kappa\times U(N)_{-\kappa}$, four complex bifundamental scalars, and $SU(4)$ R-symmetry \cite{Aharony:2008ug}.  It describes the low-energy dynamics of $N$ coincident M2-branes probing $\mathbb C^4/\mathbb Z_\kappa$.  Through gauge/gravity duality \cite{Maldacena:1997re}, its large-$N$ limit is holographically related to M-theory on $\mathrm{AdS}_4\times S^7/\mathbb Z_\kappa$ \cite{Aharony:2008ug}.  Its local operators provide a controlled arena in which gauge-theory and M-theoretic structures meet.

We focus on a highest-weight half-BPS scalar sector.  Operators of length $n$ in this sector saturate a supersymmetric bound, so their scaling dimension is protected.  They may nevertheless be assembled into many single- and multi-trace structures with identical quantum numbers.  At finite rank these structures are intrinsically nonplanar: trace relations become important and the space of nonzero operators is cut off by the gauge-group rank.  Schur operators organize this space and diagonalize its free two-point function $G^{(0)}$ \cite{Corley:2001zk,Dey:2011ea,ChakraborttyDey:2012,Caputa:2012dg}.  Restricted Schur polynomials extend this construction to multi-matrix sectors and have been used to study nonplanar ABJM dynamics \cite{BrownHeslopRamgoolam:2008,BhattacharyyaCollinsDeMelloKoch:2008,deMelloKoch:2012ABJM,deMelloKochKimMahu:2024}.  Protection does not fix the matrix of two-point coefficients, which can receive quantum corrections and induce mixing among operators.  
In ABJM the leading quantum correction to these correlators $G^{(2)}$ occurs at two loops.
This matrix by itself is not an intrinsic CFT observable: finite coupling-dependent redefinitions of protected operators change their norms and off-diagonal overlaps. 

Rather than using such operator redefinitions to perturbatively orthogonalize the protected two-point functions, we choose here to regard the leading correction as a Hermitian form on the tree-level operator space. After tree normalization, the leading correction defines the self-adjoint operator $\mathsf M^{(2)}=(G^{(0)})^{-\frac12}G^{(2)}(G^{(0)})^{-\frac12}$. The main observation of this Letter is that for gauge group $U(2)\times U(2)$, this otherwise generic spectral problem acquires an exactly solvable Young-lattice structure, with eigenvectors governed by Racah polynomials for arbitrary operator length $n$.  
For $N=2$ the Schur operators are indexed by two-row Young diagrams and form a one-dimensional chain.  After tree normalization, the mixing matrix becomes a Jacobi operator whose ground-state transform generates a reversible birth--death process with conditioned Plancherel stationary measure.  Its normal modes are Racah polynomials, and an auxiliary angular-momentum construction gives an equivalent derivation of the spectrum and eigenvectors.  At large $n$ the process approaches radial Ornstein--Uhlenbeck diffusion.  

The exact solution exposes a representation-theoretic structure before any coupling-dependent redefinitions to diagonalize the metric order by order, and supplies a distinguished basis for subsequent perturbation theory.

\section{Finite-rank two-point functions}
In manifest $SU(4)$ notation, the bifundamental scalars are $Y^A$, with conjugates $\bar Y_A$, where $A=1,\ldots,4$.  We choose the highest-weight bilinear $X=Y^1\bar Y_4$, which transforms in the adjoint of the first gauge factor.  Each factor has dimension and chosen R-charge $1/2$, so $X$ has dimension and charge one.  Operators made only from $X$ therefore saturate the BPS bound; an operator containing $n$ copies of $X$ belongs to a short multiplet with protected dimension $\Delta=n$ \cite{Dey:2011ea,Caputa:2012dg}.

The gauge-invariant trace structures of length $n$ are packaged into the Schur operators
\begin{equation}
  \OO_R(X)
  =
  \frac{1}{n!}
  \sum_{\sigma\in S_n}
  \chi_R(\sigma)\,
  \Tr\!\left(\sigma X^{\otimes n}\right),
  \qquad R\vdash n,
  \label{eq:SchurDefinition}
\end{equation}
where $\chi_R(\sigma)$ is the character of the permutation $\sigma$ in the irreducible representation $R$ of $S_n$ \cite{Fulton:1997}.  The permutation prescribes the gauge-index contractions: if its cycles have lengths $m_1,\ldots,m_p$, then $\Tr(\sigma X^{\otimes n})=\prod_{a=1}^p\Tr(X^{m_a})$.  Thus a single Schur operator is a definite linear combination of all multi-trace structures of total length $n$.

Conformal symmetry fixes the spacetime dependence of the separated-point two-point function, leaving only a numerical coefficient matrix:
\begin{equation}
  \left\langle
    \OO_R(x)\bar \OO_S(0)
  \right\rangle
  =
  \frac{G_{R,S}}{|x|^{2n}}.
  \label{eq:twoPointDefinition}
\end{equation}
We suppress this universal spacetime factor below and study the matrix $G_{R,S}$.  At tree level it is diagonal,
\begin{equation}
  G^{(0)}_{R,S}
  =
  F_R\,\delta_{RS},
  \label{eq:G0}
\end{equation}
with
\begin{equation}
  F_R=f_R^2(N), \qquad f_R(N)=\prod_{\square\in R}\bigl(N+c(\square)\bigr),
  \label{eq:fR}
\end{equation}
where a box $\square$ in row $i$ and column $j$ has content $c(\square)=j-i$ \cite{Dey:2011ea,Caputa:2012dg}.  Perturbation theory is organized in powers of the inverse Chern--Simons level $\kappa^{-1}$, and we write $G_{R,S}=G^{(0)}_{R,S}+\kappa^{-2}G^{(2)}_{R,S}+\cdots$.  The one-loop contribution at order $\kappa^{-1}$ vanishes, so the two-loop term is the first quantum correction.  We now specialize to $N=2$.  The finite-rank two-loop result derived in \cite{Bianchi:2026uts}, building on \cite{Bianchi:2024bjm,Bianchi:2025sjc}, is
\begin{equation}
  G^{(2)}_{R,S}=\cI\!\bigg[-\delta_{RS}F_R(7n+4c_R)
  \!+\!F_R F_S\!\sum_{\substack{T\vdash n-1\\T\nearrow R,\;T\nearrow S}}
  F_T^{-1}\bigg].
  \label{eq:G2general}
\end{equation}
Here $T\nearrow R$ means that $R$ is obtained from $T$ by adding one box.  The sum therefore runs over all Young diagrams $T$ with $n-1$ boxes that can be obtained by removing one box from both $R$ and $S$; each common predecessor is weighted by $F_T^{-1}$.  Moreover, $c_R=\sum_{\square\in R}c(\square)$ is the total content of $R$, and $\cI=2\zeta_2$, following \cite{Bianchi:2026uts}.

Throughout, $G^{(2)}$ is viewed as the leading deformation of the metric in the fixed Schur basis.  A finite, coupling-dependent redefinition of the operators can modify, or diagonalize, this metric order by order.  Therefore, the spectrum found below is not invariant under coupling-dependent finite operator redefinitions and should therefore not be identified with intrinsic CFT data such as anomalous dimensions.  Its significance is that the deformation in this natural representation-theoretic frame admits an exact analytic resolution, organizing the mixing and furnishing a natural leading-order eigenbasis for higher-loop perturbation theory.

At rank two, any Young diagram with three or more rows contains a vanishing factor in $f_R(2)$, so the corresponding operator has zero tree-level norm.  Hence only diagrams with at most two rows survive:
\begin{equation}
  R_k=(n-k,k),
  \qquad
  k=0,\ldots,J,
  \qquad
  J=\left\lfloor\frac n2\right\rfloor .
  \label{eq:Rk}
\end{equation}
Every operator is labeled by the single integer $k$. The full finite-rank mixing problem is defined on the ordered chain $R_0,R_1,\ldots,R_J$. Their tree-level weights read
\begin{equation}
  f_{R_k}(2)=(n-k+1)!\,k!,
  \qquad
  F_k=\bigl[(n-k+1)!\,k!\bigr]^2,
  \label{eq:Fk}
\end{equation}
while the total content is
\begin{equation}
  c_k
  =
  \frac{n(n-1)}2-k(n-k+1).
  \label{eq:ck}
\end{equation}
Two distinct two-row diagrams share a predecessor only if their labels differ by one.  Thus $G^{(2)}_{k,k'}=0$ for $|k-k'|>1$, and the finite-rank field-theory problem is tridiagonal before any further algebraic input is used.

\section{Jacobi form and birth--death process}
We introduce the tree-orthonormal operators
\begin{equation}
  \widehat{\OO}_k
  =
  \frac{\OO_{(n-k,k)}}{(n-k+1)!\,k!},
  \qquad
  \langle
  \widehat{\OO}_k
  \widehat{\OO}_{k'}^\dagger
  \rangle_{\rm tree}
  =\delta_{kk'}.
  \label{eq:Ohat}
\end{equation}
We separate the universal scalar shift in \eqref{eq:G2general} from the normalized two-loop metric correction and denote the remaining symmetric tridiagonal Jacobi operator by $H$:
\begin{equation}
  \begin{aligned}
  \mathsf M^{(2)}
  &\equiv
  (G^{(0)})^{-1/2}G^{(2)}(G^{(0)})^{-1/2}\\
  &=-\cI\left[H+7n\,\bbone\right].
  \end{aligned}
  \label{eq:MetricH}
\end{equation}
Its nearest-neighbor matrix elements are
\begin{equation}
  H_{k,k+1}=H_{k+1,k}
  =-(k+1)(n-k+1).
  \label{eq:Hoff}
\end{equation}
%An alternating-sign conjugation brings its off-diagonal entries to the conventional positive Jacobi form.
The diagonal entries are
\begin{equation}
  \begin{aligned}
  H_{kk}&=n(n-4)-1-2k(n-k+1)+\Delta_{n,k},\\
  \Delta_{n,k}&=
  \begin{cases}
    (J+1)^2, & n=2J,\ k=J,\\
    0, & \text{otherwise}.
  \end{cases}
  \end{aligned}
  \label{eq:Hdiag}
\end{equation}
The correction is present only at the rectangular endpoint $R_J=(n/2,n/2)$ for even $n$, where the box at the end of the first row is not removable.  For odd $n$, no such additional endpoint condition occurs and the generic diagonal formula applies.

We denote by
\begin{equation}
  d_R
  =
  \frac{n!}{\prod_{\square\in R}h(\square)}
\end{equation}
the dimension of the irreducible representation of $S_n$ associated with the Young diagram $R$, with $h(\square)$ the hook length \cite{Frame:1954}.  For $R_k=(n-k,k)$ this becomes
\begin{equation}
  d_k\equiv d_{(n-k,k)}
  =
  \frac{n!(n-2k+1)}
       {k!(n-k+1)!}.
  \label{eq:dk}
\end{equation}
Direct substitution shows that $d=(d_0,\ldots,d_J)^T$ satisfies
\begin{equation}
  H d=-4n\,d.
  \label{eq:ground}
\end{equation}
All components of $d$ are strictly positive, while $H$ is an irreducible real symmetric tridiagonal matrix with negative off-diagonal entries.  The Perron--Frobenius theorem \cite{HornJohnson:2012} applied to a sufficiently shifted $-H$ therefore identifies $d$ as the unique ground-state eigenvector, with energy $E_0=-4n$.

To expose the stochastic structure, we factor this profile out of a generic eigenvector and introduce the associated diagonal matrix,
\begin{equation}
  \phi_k=d_kP(k),
  \qquad
  \Psi=\operatorname{diag}(d_0,\ldots,d_J).
  \label{eq:phiP}
\end{equation}
The ground-state, or Doob, transform \cite{Doob:1957} is
\begin{equation}
  \mathcal L
  =
  -\Psi^{-1}(H+4n\,\bbone)\Psi.
  \label{eq:Doob}
\end{equation}
Eq.~\eqref{eq:ground} implies $\mathcal L\bbone=0$, so every row of $\mathcal L$ sums to zero.  Its off-diagonal entries are nonnegative and connect only neighboring states.  Hence $\mathcal L$ is the backward generator of a continuous-time birth--death process \cite{Norris:1997} on
\begin{equation}
  0\leftrightarrow1\leftrightarrow\cdots\leftrightarrow J.
  \label{eq:chain}
\end{equation}

Writing $\varepsilon=E+4n$, the eigenvalue equation $H\phi=E\phi$ becomes
\begin{align}
  (\mathcal L P) (k) &= q_k^+\bigl[P(k+1)-P(k)\bigr]
  +
  q_k^-\bigl[P(k-1)-P(k)\bigr]
  \nonumber\\
  &=-\varepsilon\,P(k),
  \label{eq:bdiff}
\end{align}
in terms of the birth and death rates
\begin{align}
  q_k^+
  &=
  (n-k+1)^2
  \frac{n-2k-1}{n-2k+1},
  \label{eq:qplus}\\
  q_k^-
  &=
  k^2
  \frac{n-2k+3}{n-2k+1}.
  \label{eq:qminus}
\end{align}
These formulas hold for $0\leq k<J$ and $1\leq k\leq J$, respectively.  The former generates the transition $k\to k+1$, while the latter generates $k\to k-1$.  At the endpoints we set \mbox{$q_0^-=q_J^+=0$}, since transitions outside the chain are forbidden.
For a birth--death chain, reversibility means that the probability flux across each edge is the same in both directions.  A stationary distribution $\pi_k$ must satisfy the detailed-balance equations
\begin{equation}
  \pi_k q_k^+
  =
  \pi_{k+1}q_{k+1}^- .
  \label{eq:DB}
\end{equation}
Substitution gives $q_k^+/q_{k+1}^-=(d_{k+1}/d_k)^2$, so the reversible weight is proportional to $d_k^2$.  After normalization, the stationary distribution is
\begin{equation}
  \pi_k
  =
  \frac{d_k^2}{C_n},
  \qquad
  C_n=\frac{1}{n+1}\binom{2n}{n},
  \label{eq:pi}
\end{equation}
where $C_n$ is the $n$th Catalan number.  The normalization follows from the identity
\begin{equation}
  \sum_{k=0}^{\lfloor n/2\rfloor}
  d_{(n-k,k)}^2
  =
  C_n.
  \label{eq:CatalanIdentity}
\end{equation}
The Plancherel measure \cite{Romik:2015} is the natural probability distribution on partitions $R\vdash n$ that assigns weight $d_R^2/n!$ to the shape $R$.  Thus $\pi_k$ is the Plancherel measure conditioned on diagrams with at most two rows.

The Catalan normalization has a direct combinatorial explanation.  The Robinson--Schensted correspondence \cite{Robinson:1938,Schensted:1961} maps each permutation to an ordered pair of standard Young tableaux of the same shape $R$.  Each shape $R$ therefore corresponds to $d_R^2$ permutations.  Schensted's theorem further states that the number of rows of $R$ equals the length of the longest decreasing subsequence of the permutation.  Summing $d_R^2$ over diagrams with at most two rows therefore counts permutations with no decreasing subsequence of length three, equivalently permutations avoiding the pattern $321$, whose number is $C_n$ \cite{SimionSchmidt:1985}, proving \eqref{eq:CatalanIdentity}.

\section{Racah solution}
The coefficients in \eqref{eq:bdiff} coincide, after identifying the parameters, with those of the second-order difference operator for Racah polynomials \cite{Koekoek:2010}.  The eigenfunctions of $\mathcal L$ on the discrete chain therefore form a finite family of Racah polynomials.  With the normalization $P_j(0)=1$, they are given by the terminating hypergeometric series
\begin{equation}
  P_j(k)
  =
  {}_4F_3\!\left(
  \begin{matrix}
    -j,\;
    j-n-\frac12,\;
    -k,\;
    k-n-1
    \\
    -n-1,\;
    \frac{1-n}{2},\;
    -\frac n2
  \end{matrix}
  ;1
  \right),
  \label{eq:Racah}
\end{equation}
with $j=0,\ldots,J$.  They obey
\begin{equation}
  \mathcal L P_j
  =
  -\varepsilon_j P_j,
  \qquad
  \varepsilon_j
  =
  2j(2n+1-2j).
  \label{eq:epsj}
\end{equation}
Consequently, the exact Hamiltonian eigensystem is
\begin{equation}
  E_j
  =
  -4n+2j(2n+1-2j), \qquad \phi_k^{(j)}=d_kP_j(k),
  \label{eq:Ej}
\end{equation}
for arbitrary operator length $n$. Therefore the exact field-theory eigenoperators in the tree-normalized basis are
\begin{equation}
  \begin{aligned}
  \widehat{\mathbb O}_j
  &=
  \frac{1}{\sqrt{C_n h_j}}
  \sum_{k=0}^{J}d_kP_j(k)\,\widehat{\OO}_k,\\
  h_j
  &=
  \sum_{k=0}^{J}\pi_kP_j(k)^2.
  \end{aligned}
  \label{eq:eigenOhat}
\end{equation}
Equivalently, in the original Schur basis and up to an overall normalization, we obtain
\begin{equation}
  \mathbb O_j
  \propto
  \sum_{k=0}^{J}
  \frac{
    n!(n-2k+1)
  }{
    (k!)^2[(n-k+1)!]^2
  }
  P_j(k)\,
  \OO_{(n-k,k)}.
  \label{eq:eigenOraw}
\end{equation}
%The proportionality sign reflects only this omitted normalization; all relative coefficients are fixed.
In the original Schur basis and using \eqref{eq:MetricH}, the generalized eigenvalues of $G^{(2)}$ relative to $G^{(0)}$, defined by $G^{(2)}v_j=\lambda_j^{(2)}G^{(0)}v_j$, are
\begin{equation}
  \lambda_j^{(2)}
  =
  -\cI
  \left[
    3n+2j(2n+1-2j)
  \right].
  \label{eq:lambdaj}
\end{equation}
Thus the finite-rank correlator is completely diagonalized through two loops for arbitrary $n$.
The spectrum is nondegenerate, so within the fixed coupling-independent Schur frame the leading deformation selects these eigenoperators uniquely up to normalization and phase.
\begin{comment}
\paragraph{Example: $n=4$}
For $n=4$ one has $J=2$ and the physical basis
\begin{equation}
  (4),\qquad (3,1),\qquad (2,2).
  \label{eq:n4basis}
\end{equation}
The general result gives
\begin{align}
  \mathbb O_0
  &\propto
  \OO_{(4)}
  +15\,\OO_{(3,1)}
  +20\,\OO_{(2,2)},
  \nonumber\\
  \mathbb O_1
  &\propto
  \OO_{(4)}
  +\OO_{(3,1)}
  -8\,\OO_{(2,2)},
  \nonumber\\
  \mathbb O_2
  &\propto
  \OO_{(4)}
  -5\,\OO_{(3,1)}
  +10\,\OO_{(2,2)},
  \label{eq:n4eigops}
\end{align}
with
\begin{equation}
  \lambda_j^{(2)}
  =
  -\cI\{12,26,32\}.
  \label{eq:n4eigvals}
\end{equation}
This three-dimensional example illustrates the general exact Racah solution valid for arbitrary $n$.
\end{comment}

\section{Auxiliary spin realization}
\label{sec:auxSU2}

Racah polynomials are closely connected with angular-momentum recoupling  \cite{Racah:1942,Varshalovich:1988}, suggesting an underlying spin realization.  Indeed, the rank-two mixing Hamiltonian admits a remarkably simple realization in terms of ordinary angular-momentum addition. 

We consider two auxiliary spins of equal magnitude
\begin{equation}
  S=\frac{n+1}{2},
\end{equation}
with generators $\bm S_1$ and $\bm S_2$, and total angular momentum
\begin{equation}
  \bm L=\bm S_1+\bm S_2.
\end{equation}
We restrict to the sector of vanishing total magnetic quantum number $M=m_1+m_2=0$ and to states that are antisymmetric under interchange of the two spins.  A convenient orthonormal basis consists of the Bell-type states
\begin{equation}
  |m\rangle_-
  =
  \frac{1}{\sqrt2}
  \left(
    |S,m\rangle\otimes|S,-m\rangle
    -
    |S,-m\rangle\otimes|S,m\rangle
  \right),
  \label{eq:antisymspinbasis}
\end{equation}
with $m>0$.  The physical two-row Young diagrams are in one-to-one correspondence with
\begin{equation}
  \begin{aligned}
    (n-k,k)&\longleftrightarrow |m_k\rangle_-,\\
    m_k&=S-k=\frac{n+1}{2}-k.
  \end{aligned}
  \label{eq:YoungSpinMap}
\end{equation}
The dimension of the antisymmetric $M=0$ sector is precisely $J+1$, matching the number of physical Schur operators.
Using
\begin{equation}
  \bm L^2
  =
  2S(S+1)
  +
  2S_{1z}S_{2z}
  +
  S_{1+}S_{2-}
  +
  S_{1-}S_{2+},
\end{equation}
we find the nearest-neighbor matrix element
\begin{equation}
  {}_-\langle m_k|
  \bm L^2
  |m_{k+1}\rangle_-
  =
  (k+1)(n-k+1).
  \label{eq:CasimirOffdiag}
\end{equation}
This is precisely minus the off-diagonal entries of the tree-normalized ABJM Hamiltonian in \eqref{eq:Hoff}.

For generic $k$ the diagonal Casimir element is
\begin{equation}
  {}_-\langle m_k|
  \bm L^2
  |m_k\rangle_-
  =
  2S(S+1)-2m_k^2.
  \label{eq:CasimirDiagGeneric}
\end{equation}
For even $n$, the last state $k=n/2$ has $m=1/2$.  In this case the transition $m=1/2\leftrightarrow -1/2$ is onto the same antisymmetric state and therefore actually produces an additional diagonal contribution
\begin{equation}
  -\left(\frac n2+1\right)^2.
  \label{eq:CasimirBoundary}
\end{equation}
This is the spin counterpart of the even-$n$ correction $\Delta_{n,k}$ in \eqref{eq:Hdiag}.  Including it, the complete rank-two Hamiltonian obeys the identity
\begin{equation}
  H
  =
  n(n-3)\,\bbone
  -
  \left.
  \bm L^2
  \right|_{M=0,\,-}.
  \label{eq:HSpinCasimir}
\end{equation}
Thus the full finite-rank mixing problem is equivalent to an ordinary two-spin Casimir restricted to a definite magnetic and exchange sector.  This representation makes the spectrum immediate.  Coupling the two spins $S=(n+1)/2$ gives $\ell=0,1,\ldots,2S=n+1$.  For two equal spins, exchange acts on the coupled state as
\begin{equation}
  P_{12}|\ell,M\rangle
  =
  (-1)^{2S-\ell}|\ell,M\rangle .
\end{equation}
Enforcing the antisymmetric sector requires $P_{12}=-1$ and $2S-\ell$ to be an odd integer, which therefore selects
\begin{equation}
  \ell_j=n-2j,
  \qquad
  j=0,\ldots,J.
  \label{eq:AllowedTotalSpin}
\end{equation}
Since $\bm L^2|\ell_j,0\rangle=\ell_j(\ell_j+1)|\ell_j,0\rangle$, Eq.~\eqref{eq:HSpinCasimir} gives
\begin{align*}
  E_j
  &=
  n(n-3)-\ell_j(\ell_j+1)
  =
  -4n+2j(2n+1-2j),
\end{align*}
reproducing exactly the Racah spectrum.

To obtain the eigenvectors, we diagonalize $\bm L^2$ in the antisymmetric $M=0$ sector and expand each coupled eigenstate $|\ell_j,0\rangle$ in the uncoupled basis $|m_k\rangle_-$, identified with the tree-orthonormal Schur basis \eqref{eq:YoungSpinMap}.  This change of basis is implemented by Clebsch--Gordan coefficients.  Up to an overall phase, the resulting components are
\begin{equation}
  u_k^{(j)}
  =
  \sqrt2\,
  \left\langle
    S,m_k;\,
    S,-m_k
    \middle|
    \ell_j,0
  \right\rangle .
  \label{eq:CGEigenvector}
\end{equation}
The prefactor $\sqrt2$ arises because the two exchanged product states have opposite Clebsch--Gordan amplitudes in the antisymmetric sector, so their projection onto the normalized difference adds the two contributions.

Thus the exact finite-rank ABJM eigenoperators may equivalently be written as
\begin{equation}
  \widehat{\mathbb O}_j
  =
  \sum_{k=0}^{J}
  u_k^{(j)}\,
  \widehat{\mathcal O}_k .
  \label{eq:CGEigenoperator}
\end{equation}
Comparison with the Racah representation above gives
\begin{equation}
  u_k^{(j)}
  =
  \pm \frac{d_k P_j(k)}{\sqrt{C_n h_j}}.
  \label{eq:CGRacahRelation}
\end{equation}

The lowest state corresponds to $j=0$ and hence to total auxiliary spin $\ell=n$.  Choosing its overall phase such that all components are positive, \eqref{eq:CGEigenvector} reduces to
\begin{equation}
  u_k^{(0)}
  =
  \frac{d_{(n-k,k)}}{\sqrt{C_n}},
  \label{eq:CGGroundState}
\end{equation}
where $C_n$ is the Catalan number.  Normalization of the coupled state therefore reproduces \eqref{eq:CatalanIdentity}, providing a spin interpretation of the Catalan normalization of the conditioned Plancherel measure.

\section{Stochastic dynamics and the large-operator limit}
The Racah solution determines the complete finite-time dynamics of the birth--death generator.  The quantity $h_j$ defined in \eqref{eq:eigenOhat} is the squared norm of $P_j$ with respect to the stationary measure $\pi_k$.  Reversibility then implies the Karlin--McGregor spectral representation \cite{KarlinMcGregor:1957}
\begin{equation}
  p_t(k,k')
  =
  \pi_{k'}
  \sum_{j=0}^{J}
  e^{-\varepsilon_j t}
  \frac{P_j(k)P_j(k')}{h_j},
  \label{eq:transitionKernel}
\end{equation}
for the transition probability from $k$ to $k'$ in time $t$.  Thus the finite-rank correlator problem defines an exactly solvable continuous-time stochastic process, rather than merely a stationary measure.

The first nonzero eigenvalue, $\varepsilon_1=4n-2$, sets the relaxation time
$\tau_{\rm rel}=\frac{1}{4n-2}$. It is therefore natural to introduce the rescaled stochastic time $s=nt$ when taking the large-$n$ limit.

A particularly simple continuum description applies near the region where the stationary measure is concentrated.  Writing the difference between the two row lengths as $r=n-2k$, the weight $d_k$ changes from increasing to decreasing when $r^2$ crosses $n+2$. Hence, dominant configurations have row imbalance of order $\sqrt n$ at large $n$.  This motivates introducing the continuum variable
\begin{equation}
  x=\frac{n-2k}{\sqrt n},
  \qquad x\ge0.
\end{equation}
The backward generator $\mathcal L$ acts on observables $f:\{0,\ldots,J\}\to\mathbb R$ on the discrete chain.  For observables that vary smoothly on the scale $\sqrt n$, we write $f(k)=\varphi(x)$.  A birth $k\to k+1$ changes $x$ by $-2/\sqrt n$, while a death $k\to k-1$ changes it by $+2/\sqrt n$.  The exact generator therefore acts as
\begin{align}
  (\mathcal L f)(k)
  ={}&
  q_k^+\!\left[\varphi\!\left(x-\frac{2}{\sqrt n}\right)-\varphi(x)\right]
  \nonumber\\
  &{}
  +
  q_k^-\!\left[\varphi\!\left(x+\frac{2}{\sqrt n}\right)-\varphi(x)\right].
\end{align}
Taylor expanding the two shifted functions and inserting the exact rates, the generator in the rescaled time $s=nt$ is $\mathcal L/n$ and converges to
\begin{equation}
  \frac{1}{n}\mathcal L
  \;\longrightarrow\;
  \mathcal G
  =
  \frac{d^2}{dx^2}
  +
  2\left(\frac1x-x\right)\frac{d}{dx}
  \label{eq:radialOU}
\end{equation}
for fixed $x>0$, at leading order in $1/\sqrt n$.  This is the radial Ornstein--Uhlenbeck generator in three dimensions \cite{UhlenbeckOrnstein:1930,KarlinTaylor:1981}.  Correspondingly, the conditioned Plancherel measure approaches the Maxwell distribution, the probability law for the radius of a three-dimensional isotropic Gaussian \cite{Sniady:2007,Matsumoto:2008},
\begin{equation}
  \frac{\pi_k}{\Delta x}
  \;\longrightarrow\;
  \rho_\infty(x)
  =
  \frac{4}{\sqrt{\pi}}\,
  x^2 e^{-x^2},
  \qquad x\geq0,
  \label{eq:MaxwellLimit}
\end{equation}
where $\Delta x=2/\sqrt n$ is the lattice spacing.  In other words, the Catalan-normalized probability measure on two-row Young diagrams has a universal Gaussian radial limit.

Expanding the terminating Racah representation at fixed mode number $j$ and fixed $x$, one finds 
\begin{equation}
  \frac{(-1)^j n^j}{j!}P_j(k)
  \;\longrightarrow\;
  L_j^{(1/2)}(x^2)
  \label{eq:RacahLaguerre}
\end{equation}
where $L_j^{(1/2)}$ is a generalized Laguerre polynomial \cite{Koekoek:2010}.  Indeed,
\begin{equation}
  \mathcal G
  L_j^{(1/2)}(x^2)
  =
  -4j\,
  L_j^{(1/2)}(x^2),
  \label{eq:LaguerreEigen}
\end{equation}
in agreement with the exact discrete result for fixed $j$,
\begin{equation}
  \frac{\varepsilon_j}{n}
  =
  \frac{2j(2n+1-2j)}{n}
  \longrightarrow
  4j.
\end{equation}

\section{Conclusions}
These results reveal a direct and unexpected chain of equivalences.  In the fixed, coupling-independent Schur frame, the leading deformation of the protected two-point metric in $U(2)\times U(2)$ ABJM theory reduces to a Jacobi operator on a chain of two-row Young diagrams.  A ground-state transform maps this operator to the generator of a reversible birth--death process, whose normal modes are Racah polynomials.  Their orthogonality resolves the deformation relative to the tree-level metric, while their eigenvalues characterize the two-loop deformation in this fixed coupling-independent frame.  Protected-operator normalizations are not themselves intrinsic CFT data and may be altered by finite coupling-dependent redefinitions; the nontrivial result is the closed algebraic structure displayed before any such redefinition, identifying the principal directions of the leading quantum deformation of the tree-level metric.

The rank-two solution is special because the finite-rank constraint leaves only the two-row diagrams $R_k=(n-k,k)$: each $k$ labels a single state, and removing or adding a box connects only neighboring values of $k$.  The mixing problem therefore reduces to a scalar three-term recurrence on a chain.  At higher rank, diagrams with more rows survive and the Young-diagram graph branches, so several independent mixing channels can occur and one expects a block- or multivariate recurrence rather than the scalar three-term one.  Nevertheless, the rank-two problem suggests that finite-rank protected correlators may admit a broader formulation in terms of weighted Young-lattice dynamics, ladder algebras and multivariate discrete orthogonal systems, which will be presented elsewhere.  
The Racah basis also provides a distinguished starting point for higher-loop perturbation theory where it would be interesting to investigate the emergence of unexpected algebraic structure underlying the protected two-point functions. It would also be interesting to determine whether the Racah basis simplifies protected three-point functions and OPE coefficients \cite{HiranoKristjansenYoung:2012,Young:2014Chiral,Young:2014Extremal,Bianchi:2020ThreePoint}, which could provide a more intrinsic characterization of the representation-theoretic structure uncovered here.

\bigskip

\acknowledgments

This work was supported by Fondo Nacional de Desarrollo Científico y Tecnológico, through Fondecyt Exploración 13250014.
%\end{acknowledgments}

\bibliography{references}

\end{document}